\documentclass[twocolumn,twoside,nofootinbib,showpacs,prd,aps,tightenlines,10pt]{revtex4-1}

\usepackage{graphicx}
\usepackage{amsmath}
\usepackage{amssymb}
\usepackage{slashed}
\usepackage[hyperfootnotes=false,bookmarks=false]{hyperref}

\newcommand{\eq}[1]{Eq.~\eqref{#1}}
\newcommand{\eqs}[2]{Eqs.~\eqref{#1} and \eqref{#2}}
\renewcommand{\sec}[1]{Sec.~\ref{sec:#1}}

\newcommand{\fig}[1]{Fig.~\ref{fig:#1}}

\def\nn{\nonumber \\}
\def\vq{Q}
\def\lQ{\log \frac{Q^2}{\mu^2}}
\def\lM{\log \frac{M^2}{\mu^2}}

\def\bn{\bar n}
\def\rd{{\rm d}}
\def\vev#1{\left\langle #1 \right \rangle}

\def\sdt{\!\cdot\!}
\def\bT{\mathbf{T}}
\def\op{{\mathcal{O}}}
\def\lra{\leftrightarrow}

\def\al{\alpha}
\def\ga{\gamma}
\def\de{\delta}
\def\De{\Delta}
\def\eps{\epsilon}
\def\vp{\varphi}
\def\zb{\textrm{\o}}

\begin{document}

%%%%%%%%%%%%%%%%%%%%%%%%%%%%%%%%%%%%%%%%%%%%%%%%%%%%%%%%%%%%%%%%%%%%%%%%%%%%%%%%
% title page
%%%%%%%%%%%%%%%%%%%%%%%%%%%%%%%%%%%%%%%%%%%%%%%%%%%%%%%%%%%%%%%%%%%%%%%%%%%%%%%%

\title{Electroweak Radiative Corrections to Higgs Production via Vector Boson Fusion using Soft-Collinear Effective Theory}

\author{Andreas Fuhrer}
\author{Aneesh V.~Manohar}
\author{Wouter J.~Waalewijn}
\affiliation{Department of Physics, University of California at San Diego,
  La Jolla, CA 92093\vspace{2ex} }

%%%%%%%%%%%%%%%%%%%%%%%%%%%%%%%%%%%%%%%%%%%%%%%%%%%%%%%%%%%%%%%%%%%%%%%%%%%%%%%%
\begin{abstract}
Soft-collinear effective theory (SCET) is applied to compute electroweak radiative corrections to Higgs production via gauge boson fusion, $q q \to q q H$. There are several novel features which make this process an interesting application of SCET: The amplitude is proportional to the Higgs vacuum expectation value (VEV), and so is not a gauge singlet amplitude. Standard resummation methods require a gauge singlet operator and do not apply here. The SCET analysis requires operators with both collinear and soft external fields, with the Higgs VEV being described by an external soft $\phi$ field. There is a scalar soft-collinear transition operator in the SCET Lagrangian which contributes to the scattering amplitude, and is derived here.
\end{abstract}
%%%%%%%%%%%%%%%%%%%%%%%%%%%%%%%%%%%%%%%%%%%%%%%%%%%%%%%%%%%%%%%%%%%%%%%%%%%%%%%%


\maketitle

%%%%%%%%%%%%%%%%%%%%%%%%%%%%%%%%%%%%%%%%%%%%%%%%%%%%%%%%%%%%%%%%%%%%%%%%%%%%%%%%
\section{Introduction}\label{sec:intro}
%%%%%%%%%%%%%%%%%%%%%%%%%%%%%%%%%%%%%%%%%%%%%%%%%%%%%%%%%%%%%%%%%%%%%%%%%%%%%%%%

High energy scattering processes at partonic center-of-mass energy $\sqrt{\hat s}$ have large radiative electroweak corrections due to the presence of $\alpha^n\log^{2n} \hat s/M_Z^2$ terms at $n^{\rm th}$ order in perturbation theory, which lead to a breakdown of fixed-order perturbation theory for large values of $\hat s$. These large radiative corrections can be systematically computed by various resummation methods~\cite{Sterman:1995fz}. One such method uses Soft-Collinear Effective Theory (SCET)~\cite{BFL,SCET1,SCET2,BPS}, summing the large logarithms by the renormalization group evolution in the effective theory. This effective field theory (EFT) approach for spontaneously broken gauge theories such as the $SU(2) \times U(1)$ electroweak theory was developed in Refs.~\cite{Chiu:2007yn,Chiu:2007dg,Chiu:2008vv,Chiu:2009mg,Chiu:2009ft}. The purely electroweak corrections at LHC energies are significant (e.g.~the electroweak corrections to transverse $W$ pair production are 37\% at $\sqrt{\hat s} = $2~TeV) and need to be included to get reliable cross-section predictions. 

In this paper, we apply effective theory methods to study electroweak radiative corrections to an important LHC process, Higgs production via vector boson fusion (VBF). QCD corrections to the VBF process have been calculated at next-to-leading order (NLO) for the inclusive cross section in Ref.~\cite{Han:1992hr} and for the differential cross section in Refs.~\cite{Figy:2003nv,Berger:2004pca,Arnold:2008rz}. The gluon-induced contribution at next-to-next-to-leading order (NNLO) was determined in Ref.~\cite{Harlander:2008xn} and the structure-function approach was employed in Refs.~\cite{Bolzoni:2010as,Bolzoni:2010xr} to get an accurate approximation to the full NNLO QCD corrections. The NLO electroweak corrections were determined in Refs.~\cite{Ciccolini:2007jr,Ciccolini:2007ec}, which are comparable in size to the QCD corrections and thus numerically important. The EFT method provides additional insight into the form of the radiative corrections. The computation has some interesting field theory features which are discussed below. 

\begin{figure}
\includegraphics[width=4cm]{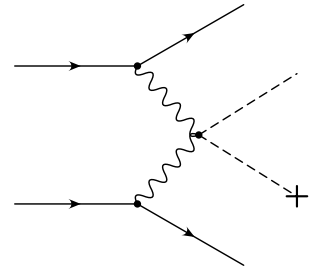}
\caption{\label{fig:fusion} Higgs production via gauge boson fusion. The solid lines are fermions, wiggly lines are gauge bosons, dashed lines are scalars, and ``$+$" denotes a Higgs VEV.}
\end{figure}

\begin{figure}
\includegraphics[width=4cm]{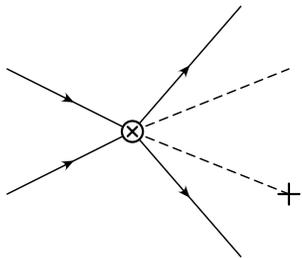}
\caption{\label{fig:fusioneft} Higgs production via gauge boson fusion, with the gauge bosons integrated out. Solid lines are fermions, dashed lines are scalars, and ``$+$" denotes a Higgs VEV.}
\end{figure}

The basic diagram is shown in Fig.~\ref{fig:fusion}.
%%%
%%%
We will compute the radiative corrections to this process in the hard-scattering regime, where the outgoing fermions are not collinear to the incoming fermions, so that the gauge bosons are off-shell, and Fig.~\ref{fig:fusion} reduces to the effective theory diagram in Fig.~\ref{fig:fusioneft}. The extension to the case where one or both gauge bosons are nearly on-shell is discussed at the end of the paper. There is an important new feature which make the vector boson process particularly interesting --- the amplitude is explicitly proportional to the $SU(2) \times U(1)$ symmetry breaking vacuum expectation value (VEV) of the Higgs field $v$,\footnote{The standard model Lagrangian has a $\phi \to -\phi$ symmetry if the Yukawa couplings are neglected, which would not allow for single Higgs production. This symmetry is broken by the VEV $v$, and the single Higgs amplitude is proportional to $v$.} so that the EFT operator is not a gauge singlet. Gauge invariance of the hard scattering amplitude plays a key role in the development of standard resummation formul\ae~\cite{Sterman:1995fz}, which can thus no longer be used for this problem. The QCD corrections do not suffer this complication, since they receive no contribution from the Higgs sector.
%%%
%%%

In the method developed in Refs.~\cite{Chiu:2007dg,Chiu:2007yn,Chiu:2009mg,Chiu:2008vv,Chiu:2009ft}, one first matches onto SCET at a high scale $\mu_h \simeq \sqrt{\hat s}$. This matching can be done in the unbroken $SU(3) \times SU(2) \times U(1)$ gauge theory, since symmetry breaking effects are proportional to $v$, and thus are suppressed by powers of $v/\sqrt{\hat s}$. One then runs the effective theory operators to a low scale $\mu_l \simeq M_Z$ using the SCET renormalization group evolution, which sums the $\alpha^n \log^{k\le 2n} \hat s/M_Z^2$ terms. At the low-scale, one integrates out the $W$ and $Z$, and matches onto a $SU(3) \times U(1)$ effective theory, which only contains gluons and photons. The effects of $SU(2) \times U(1)$ symmetry breaking enter into this low-scale matching. The high-scale matching, and running between the high scale $\mu_h \sim \sqrt{\hat s}$ and the low scale $\mu_l \sim M_Z$ can be done using $SU(3) \times SU(2) \times U(1)$ invariant operators in the unbroken theory.

In our scattering process, the incoming and outgoing fermions, and outgoing Higgs particle are energetic, and are treated as collinear fields in SCET. There is also a scalar field $\phi$ that is replaced by the Higgs VEV $\vev{\phi}=v/\sqrt{2}$, and so carries zero four-momentum. The Higgs VEV can be included by matching at the high scale onto gauge-invariant SCET operators with an external soft Higgs field, which is replaced by the Higgs VEV at the low scale $\mu_l$. The SCET operators contain both collinear and soft external fields, and the formalism for including external soft matter fields will be developed in this paper. There is a gauge-invariant soft $\leftrightarrow$ collinear transition operator in the subleading SCET Lagrangian that contributes to the vector boson fusion amplitude, and is derived in Sec.~\ref{sec:eft} using tree-level matching. The Lagrangian of Sec.~\ref{sec:eft} is used in Sec.~\ref{sec:oneloop} to compute the one-loop form factor for the $\phi^\dagger \phi$ operator. The formalism is then applied to the vector boson fusion amplitude in Sec.~\ref{sec:vbf}. The scattering amplitude for energetic external particles (i.e.\ with collinear external fields) was written in Refs.~\cite{Chiu:2009mg,Chiu:2009ft} in terms of process-independent collinear functions, and a soft-function with a universal form. The amplitude for gauge boson fusion can be written in the same way for the fermionic part, with an additional contribution from the Higgs sector.  Some subtleties in evaluating SCET integrals using contour integration are discussed in Appendix~\ref{sec:app}. The list of SCET operators for vector boson fusion is given in Appendix~\ref{sec:basis}.

\medskip

\noindent {\sl Notation:}
We take all momenta to be incoming, except for the Sudakov case discussed in \sec{oneloop}. The light-cone reference vectors are chosen to be
%%%
\begin{eqnarray}
n_i^\mu = \pm(1,\vec n_i)\,, \quad
\bn_i^\mu = \pm(1,-\vec n_i)\,,
\end{eqnarray}
%%%
where $\vec n_i$ points in the direction of particle $i$. The plus (minus) sign occurs if particle $i$ is incoming (outgoing), except in the Sudakov case where we will always take the plus sign.
The group theory generators will be treated as operators $\mathbf{T}^a_i$, which act on particle $j$ as,
%%%
\begin{eqnarray}
\big(\mathbf{T}^a_i \psi_j\big)_\alpha &=& -T^a_{\alpha \beta} \psi_{j\beta}\, \delta_{ij}\,, \nn
\big(\mathbf{T}^a_i \bar \psi_j\big)_\alpha &=& \bar \psi_{j\beta} T^a_{\beta \alpha}\, \delta_{ij}\,.
\end{eqnarray}
%%%
The covariant derivative is $iD=i\partial - g A$. The scalar multiplet is
%%%
\begin{eqnarray} \label{goldst}
\phi &=& \frac{1}{\sqrt 2}\left( \begin{array}{c}
\varphi^2 + i \varphi^1 \\
v + h - i \varphi_3 \end{array} \right)\,,
\end{eqnarray}
%%%
where $\varphi^a$ are the unphysical Goldstone bosons, and $h$ is the Higgs. The $\lambda \phi^4$ coupling is normalized so that the Higgs mass is $M_h^2=2 \lambda v^2$, with $v \sim 246$~GeV.

%%%%%%%%%%%%%%%%%%%%%%%%%%%%%%%%%%%%%%%%%%%%%%%%%%%%%%%%%%%%%%%%%%%%%%%%%%%%%%%%
\section{The Effective Theory Lagrangian}\label{sec:eft}
%%%%%%%%%%%%%%%%%%%%%%%%%%%%%%%%%%%%%%%%%%%%%%%%%%%%%%%%%%%%%%%%%%%%%%%%%%%%%%%%

In this section, we review the standard form of the SCET operators for processes involving external collinear fields, and then extend the results to operators involving external soft and collinear fields. We will compute the tree-level matching explicitly, and show how the Wilson lines contained in the operators are built up. The subleading SCET Lagrangian contains transition operators between soft and collinear fields, which we derive from this tree-level matching calculation, and will need to study gauge boson fusion.

The effective theory discussed here is ${\rm SCET}_{\rm EW}$ studied in Refs.~\cite{Chiu:2007yn,Chiu:2007dg,Chiu:2008vv,Chiu:2009mg,Chiu:2009ft}. The $n_i$-collinear fields describe particles with momentum $p_i$ near the $n_i$ direction, which scale like $\bn_i \cdot p_i \sim \sqrt{\hat s}$, $n_i \cdot p_i \sim \lambda^2 \sqrt{\hat s}$ and $p_{i\perp} \sim \lambda \sqrt{s}$. Here $\lambda \sim M/\sqrt{\hat s} \ll 1$ is the SCET power counting parameter (not to be confused with the Higgs self-coupling), and $M$ is the gauge boson mass. There is also a mass-mode field, which contains massive gauge bosons with all momentum components scaling like $\lambda \sqrt{\hat s} \sim M$. To describe the VEV in the unbroken phase of the gauge theory, we include a soft scalar field with momentum $p_0^\mu \sim \lambda^2 \sqrt{\hat s}$.

Consider first a hard-scattering process with $r$ external energetic particles, in directions $n_i$, $i=1,\ldots,r$. The effective interaction can schematically be represented as an operator
%%%
\begin{eqnarray}
\prod_i \psi_i\,,
\label{eq2}
\end{eqnarray}
%%%
where all the fields are treated as incoming for notational simplicity, and $\psi$ represents fermions or scalars. The gauge indices on the full theory fields $\psi_i$ are combined to form a gauge-invariant operator.
%%%
\begin{figure}
\includegraphics[width=2.5cm]{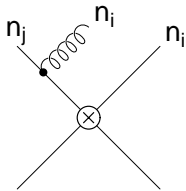}
\caption{\label{fig:wilson} Collinear gauge boson interactions which generate the collinear Wilson lines.}
\end{figure}
%%%
At the high scale $\mu_h \sim \sqrt{\hat s}$, the operator in Eq.~(\ref{eq2}) is matched onto an SCET operator with $r$ external collinear fields,
%%%
\begin{eqnarray}
\prod_i  (W^\dagger_{n_i} \xi_{n_i}) = \prod_i   \chi_{n_i}, \qquad \chi_{n_i} \equiv W^\dagger_{n_i} \xi_{n_i} \,,
\label{eq3}
\end{eqnarray}
%%%
where $\xi_{n_i}$ is a collinear field for energetic particles moving in the direction $n_i$, and $W_{n_i}$ is a $n_i$-collinear Wilson line. The combination $\chi_{n_i}=W^\dagger_{n_i} \xi_{n_i}$ is gauge invariant under $n_i$-collinear gauge transformations. 

It is instructive to see the origin of the collinear Wilson lines. If $n_i$-collinear gauge bosons interact with energetic particles moving in the $n_j$ direction (see Fig.~\ref{fig:wilson}), the intermediate particles are off shell by an amount of order $\sqrt{\hat s}$, and can be integrated out. The resulting operator is a Wilson line involving $n_i$-collinear gauge bosons, transforming according to the representation $\mathfrak{R}_j$ of particle $j$. The $n_i$-collinear Wilson lines for all particles $n_j$, $j=1,\ldots, r$ with $j \not =i$ are combined into a $n_i$-collinear Wilson line transforming as the complex conjugate representation of particle $i$, $\overline{\mathfrak{R}}_i$. This is because the gauge indices of the operator in Eq.~(\ref{eq2}) are combined to form a gauge singlet, so the tensor product $\otimes_{j\not=i} \mathfrak{R}_j$ of the Wilson lines combine into a single Wilson line transforming as $\overline{\mathfrak{R}}_i$. This allows the interaction to be written as in Eq.~(\ref{eq3}), with a single Wilson line  $W^\dagger_{n_i}$ for each collinear direction $n_i$, transforming in the representation $\overline{\mathfrak{R}}_i$ conjugate to particle $i$.\footnote{There are important subtleties which arise because of the regulator used in SCET, which are discussed in Refs.~\cite{Chiu:2009mg,Chiu:2009yx,zerobin}.}

%%%
\begin{figure}
\includegraphics[width=2.5cm]{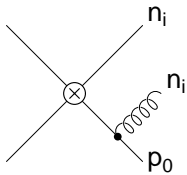}
\caption{\label{fig:p0} Collinear gauge boson interactions with a soft field, which generate the collinear Wilson lines and the subleading interaction in \eq{eq7}.}
\end{figure}
%%%
For the Higgs production amplitude of Fig.~\ref{fig:fusioneft}, we need an interaction of the form Eq.~(\ref{eq2}), where one of the scalar fields has a momentum which is soft. We will label the soft scalar field by index $0$, and the collinear fields by index $i$, $i=1,\ldots,r-1$. 
We now perform the matching from the full theory onto SCET analogous to the matching of \eq{eq2} onto \eq{eq3} in the case where all the particles are collinear. The emission of an $n_i$-collinear gauge boson from the $n_j$-collinear particle produces a $n_i$-collinear Wilson line transforming under the representation $\mathfrak{R}_j$, as before. However, these do not combine into a single Wilson line transforming as $\overline{\mathfrak{R}}_i$, because the soft field is not a gauge singlet.

%%%
\begin{figure}
\includegraphics[width=2cm]{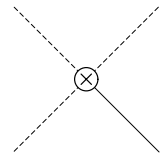}
\caption{\label{fig:veft}  SCET vertex corresponding to Eq.~(\ref{eq6}). The dashed lines are collinear fields, and the solid line is a soft field.}
\end{figure}
%%%
There are of course also graphs where the $n_i$ collinear gauge boson is emitted from the soft line, as shown in Fig.~\ref{fig:p0}. The one-gauge boson emission graph in Fig.~\ref{fig:p0} is, for an external soft scalar,
proportional to
%%%
\begin{eqnarray}
\hspace{-3ex}
\frac{(2p_0 \!+\! k)\sdt \varepsilon}{(p_0\!+\!k)^2-m^2}
&=& \frac{\bn_i \sdt \varepsilon}{\bn_i \sdt k}
+\left[\frac{(2p_0\!+\!k) \sdt \varepsilon}{(p_0\!+\!k)^2-m^2} - \frac{\bn_i \sdt \varepsilon}{\bn_i \sdt k}\right] ,
\label{eq4}
\end{eqnarray}
%%%
where $p_0$ and the gauge boson momentum $k$ are incoming, $p_0$ is soft and $k$ is $n_i$ collinear, $\varepsilon$ is the gauge boson polarization and $m$ is the mass of the scalar multiplet $\phi$. The first term generates a $n_i$-collinear Wilson line transforming as the representation $\mathfrak{R}_0$ of the soft particle. Combined with the $n_i$-collinear Wilson lines generated by gauge boson emission from the collinear lines $n_j$ ($j\not=i$), this produce a single Wilson line transforming as $\overline{\mathfrak{R}}_i$. This leads to an EFT vertex of the form
%%%
\begin{eqnarray}
\psi_0 \prod_{i=1}^{r-1}  (W^\dagger_{n_i} \xi_{n_i}) = \psi_0 \prod_{i=1}^{r-1}  \chi_{n_i}\,,
\label{eq6}
\end{eqnarray}
%%%
which can be represented graphically as in Fig.~\ref{fig:veft}.

%%%
\begin{figure}
\includegraphics[width=3cm]{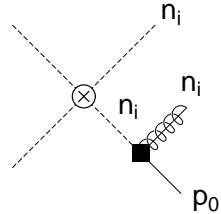}
\caption{\label{fig:neft}  Graph involving the soft-collinear transition vertex, denoted by a ``$\blacksquare$". The dashed lines are collinear fields, and the solid line is a soft field.}
\end{figure}
%%%

\begin{figure}
\includegraphics[width=3cm]{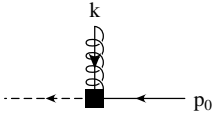}
\caption{\label{fig:neft2} The soft-collinear transition vertex.}
\end{figure}

The remaining terms in \eq{eq4} reduce to
%%%
\begin{eqnarray}
\frac{1}{(p_0+k)^2-m^2} \frac{1}{\bn_i \sdt k} \left[ (k \sdt \varepsilon) ( \bn_i \sdt k ) - (\bn_i \sdt \varepsilon) k^2\right] ,
\label{eq5}
\end{eqnarray}
%%%
to leading order in the SCET power counting. This additional contribution is given by a new SCET diagram shown in  Fig.~\ref{fig:neft}.
The new vertex ``$\blacksquare$", shown in Fig.~\ref{fig:neft2}, arises from
%%%
%%%
a gauge-invariant term in the subleading SCET Lagrangian
%%%
\begin{eqnarray}
L &=& -ig \sum \left(\xi^\dagger_n \frac{ \bn_\mu \Pi_{n,\nu} }{\bn \sdt \Pi_n} F_{n}^{\mu \nu} W_n \right)\psi_0
+\text{h.c.}\,.
\label{eq7}
\end{eqnarray}
%%%
Here the collinear covariant derivative $\Pi_n$ is given by $\Pi_n^\mu = \mathcal{P}_n^\mu - g A_n^\mu$, $\mathcal{P}_n$ is the label momentum operator and $A_n^\mu$ and $F_n^{\mu \nu}$ are the $n$-collinear gauge field and field strength. [Note that $n \cdot \Pi_n$ does not appear in \eq{eq7}.]
\eq{eq7} gives the Feynman rule
%%%
\begin{eqnarray}
-ig T^a \frac{1}{\bn \sdt k} \left[ ( \bn \sdt k )k^\alpha  -k^2 \bn^\alpha \right] \,,
\label{eq110}
\end{eqnarray}
%%%
for one-gauge boson emission. The additional factor $1/[(p_0+k)^2-m^2]$ needed to get Eq.~(\ref{eq5}) arises from the intermediate collinear propagator. 
The operator at the hard vertex ``$\otimes$" in Fig.~\ref{fig:neft} has two collinear fields in the $n_i$ direction, one of which gets turned into a soft field by the new vertex. This soft $\leftrightarrow$ collinear transition vertex does not contribute to the computations in Refs.~\cite{Chiu:2009mg,Chiu:2009ft}, which only looked at processes with external collinear particles in different directions.

In terms of the SCET $B$ field,
%%%
\begin{eqnarray}
B_n^\nu &=& \frac{i}{\bar n \sdt \Pi_n}\bar n_\mu F_n^{\mu \nu}\,,
\end{eqnarray}
%%%
Eq.~(\ref{eq7}) is\footnote{One may remove the $n \sdt B_n$ dependence in \eq{eq10}, by using the equations of motion, to write the Lagrangian in terms of standard building blocks, $\xi_n$, $\psi_0$ and $B_{n\perp}$ \cite{Marcantonini:2008qn}. This removes the single gluon emissions from the subleading Lagrangian and introduces a $\xi_n^\dagger \xi_n \xi_n^\dagger \psi_0$ contact interaction. This field redefinition changes the diagrams in \sec{oneloop}, but the result for the total on-shell scattering amplitude is unaffected.}
%%%
\begin{eqnarray}
L &=& -g \sum \left(\xi^\dagger_n \Pi_{n,\nu}  B_{n}^{\nu} W_n\right)\psi_0 +\text{h.c.}\,.
\label{eq10}
\end{eqnarray}
%%%
The (unspecified) sums in \eqs{eq7}{eq10} run over all possible values of the label momenta, subject to momentum conservation, and over all possible collinear directions $n$. The label momenta have to be nonzero on collinear fields to avoid double counting the soft degrees of freedom, so the label sum excludes the zero-bin~\cite{zerobin}. This will be important later on. 

If the soft particle is a fermion, the corresponding subleading SCET Lagrangian is
%%%
\begin{eqnarray}
\hspace{-3ex}
L &=& -ig\left(\bar \xi_n \frac{ \bn_\mu \gamma_\nu^\perp }{\bn \sdt \Pi} F_n^{\mu \nu}W_n\right)\psi_0 + \text{h.c.} \nn
 &=& -g \left(\bar \xi_n \slashed{B}_n^\perp W_n \right)\psi_0 + \text{h.c.}\,.
\label{eq10a}
\end{eqnarray}
%%%
The case of soft fermions has been extensively studied in the literature~\cite{Chay:2002vy,Beneke:2002ph,Feldmann:2002cm}. The interaction Eq.~(\ref{eq10a}) has been obtained previously~\cite{Pirjol:2002km} using reparametrization invariance~\cite{Luke:1992cs,Manohar:2002fd}.

More transition operators are generated at higher order in perturbation theory, but will not be required for our analysis as they are higher order in the power counting. For example, at order $g^2$ for soft scalar particles, there is also the interaction term
%%%
\begin{eqnarray}
L &=&  g^2 \sum (\xi^\dagger_n B_{n,\perp \alpha}B_{n,\perp}^\alpha W_n)\psi_0\,,
\end{eqnarray}
%%%
which can be obtained by matching the two-gauge boson amplitude onto the effective theory.

We conclude this section with a brief discussion of SCET in the broken phase. An important difference between the broken and unbroken phases is in the choice of gauge-fixing term. While the Lagrangian depends on the gauge-fixing term, the $S$-matrix does not.

We start by deriving SCET in the broken phase from the full theory in the broken phase. In the unbroken theory we used Feynman gauge with gauge fixing term
%%%
\begin{eqnarray}
-\frac 12 \left(\partial^\mu A^a_\mu\right)^2\,,
\end{eqnarray}
%%%
whereas in the broken theory we will use $R_{\xi=1}$ gauge with gauge fixing term
\begin{eqnarray}
\label{gaugefix}
-\frac 12 \left(\partial^\mu A^a_\mu + M \varphi^a\right)^2\,.
\end{eqnarray}
Here $\varphi^a$ are the Goldstone bosons in \eq{goldst} and $M$ is the gauge boson mass. The $A\varphi$ cross-terms from the $R_\xi$ gauge-fixing term cancel the $A\varphi$ mixing terms in the kinetic energy of the scalar, and the Goldstone bosons obtain a mass $M$. In Feynman gauge, the gauge boson can couple the VEV to an intermediate $\varphi$ particle, but such a coupling is absent in the $R_\xi$ gauge. Thus, there are no emissions from the line associated with the VEV in $R_\xi$ gauge. In particular, we do not get the contribution in the $\mathfrak{R}_0$ representation to the $n_i$-collinear Wilson line, which was present in the unbroken phase.

Alternatively, we may obtain SCET in the broken phase from SCET in the unbroken phase by changing the gauge-fixing term, which should lead to the same result.  For simplicity, we only show that the contributions to the emission of a single gauge boson from the line associated with the VEV cancel. Adding up the contribution from the Wilson line $W_n$, the contribution from the subleading Lagrangian in \eq{eq7}, where $\psi_0$ has been replaced by its VEV, and the contribution from the gauge-fixing term in \eq{gaugefix}, we get
%%%
\begin{eqnarray} \label{brcancel}
 &&\frac{\bn^\al}{\bn \sdt k} + \frac{1}{k^2-M^2}
 \frac{1}{\bn \sdt k} \left[ ( \bn \sdt k )k^\alpha  -k^2 \bn^\alpha \right] - \frac{1}{k^2-M^2} k^\al 
 \nn &&\quad
 = -\frac{1}{k^2-M^2} \frac{1}{\bn \sdt k} M^2 \bn^\al \,,
\end{eqnarray}
%%%
where we have left out an overall factor of $M = g v/2$. The terms do not seem to cancel. However, there is another contribution due to a subtlety when switching to the broken phase. In the unbroken phase, the soft field is of order $\psi_0 \sim \lambda^2$, where $\lambda$ is the power counting parameter (not to be confused with the Higgs self-coupling). In the broken phase we get $\langle \psi_0 \rangle \sim M \sim \lambda$, so its power counting changes. This implies that subleading terms in the unbroken Lagrangian can get promoted to leading order, when switching to the broken phase. Using  $\phi \to \xi_n + W_n \psi_0$, decomposing $\phi$ and $\xi_n$ as in \eq{goldst}, we obtain 
%%%
\begin{eqnarray}
\hspace{-5ex}
  - \frac 12 M^2 \varphi^a \varphi^a = - \frac 12 M^2 \left(\varphi_n^a \varphi_n^a + i v \frac{g \bn \sdt A_n}{\bn \sdt k} \varphi_n^a + \dots \right)\!,
\end{eqnarray}
%%%
from the Goldstone boson mass term by emitting a gauge boson from the Wilson line. The second term on the right hand side is a subleading term that gets promoted to leading order when $\phi_0$ gets a VEV. It gives the missing contribution necessary to cancel  \eq{brcancel}.

%%%%%%%%%%%%%%%%%%%%%%%%%%%%%%%%%%%%%%%%%%%%%%%%%%%%%%%%%%%%%%%%%%%%%%%%%%%%%%%%
\section{One Loop}\label{sec:oneloop}
%%%%%%%%%%%%%%%%%%%%%%%%%%%%%%%%%%%%%%%%%%%%%%%%%%%%%%%%%%%%%%%%%%%%%%%%%%%%%%%%

%%%
\begin{figure}
\includegraphics[width=3cm]{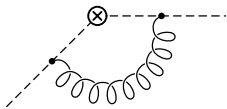}
\caption{\label{fig:oneloop} One-loop scattering of a scalar field by the external operator $\phi^\dagger \phi$.}
\end{figure}
%%%

A nontrivial check of the results of the previous section is to verify that the effective theory correctly reproduces the infrared structure of the full theory at the one-loop level. The process we will consider is shown in Fig.~\ref{fig:oneloop}: the one-loop scattering of a scalar field by the external operator $\phi^\dagger \phi$. The incoming scalar momentum $p_0$ is taken to be soft, the outgoing scalar momentum $p_f$ is taken to be collinear and we will assume for simplicity that the external particles are massless. The general scattering process of this kind, such as vector boson fusion in Fig.~\ref{fig:fusioneft}, has several collinear particles and one soft particle. There, the one-loop vertex graphs are a sum of graphs connecting two collinear particles, or a collinear particle with a soft particle. We already know that the infrared behavior of the collinear-collinear graphs in the full and effective theories agree. If the effective theory correctly reproduces Fig.~\ref{fig:oneloop}, then it will also correctly reproduce the collinear-soft graphs, and hence the total one-loop correction. We will apply the results derived here to vector boson fusion in \sec{vbf}. 

The vertex graph depends on $p_0 \cdot p_f$ and the gauge boson mass $M^2$. Since $p_f$ is collinear and $p_0$ is soft, $p_0 \cdot p_f \sim \lambda^2 \hat s$, where $\lambda$ is the SCET power counting parameter. The SCET power counting for the massive gauge theory is such that $\lambda^2 \sim M^2/\hat s$, so that $p_0 \cdot p_f \sim M^2$. This scale is not a hard scale and does not get integrated out, so the effective theory must reproduce the entire $p_0 \cdot p_f/M^2$ dependence of the vertex graph. Eventually, we are interested in the case $p_0 \to 0$, since the soft field will turn into the Higgs VEV. The on-shell vertex graph in the full theory is
%%%
\begin{eqnarray}
D_F &=&\frac{\alpha}{4\pi} \bT_0 \sdt \bT_f\, I_F\,, \nn
I_F &=&i (4\pi)^2 f_\epsilon \int\! \frac{{\rm d}^d k}{(2 \pi )^d} \frac{(2p_0+k) \sdt (2p_f+k)}{(k+p_0)^2(k+p_f)^2(k^2-M^2)}\nn
&=& -\frac{1}{\epsilon} - \lQ  + 2 \lM \nn
&& + \left(2 -\frac{M^2}{\vq^2}\right) \left[ \frac{\pi^2}{6}-\text{Li}_2\left(1-\frac{\vq^2-i0^+}{M^2}\right)\right]\!,
 \label{eq39}
\end{eqnarray} 
%%%
with $\vq^2=-( p_0 - p_f)^2$ and $f_\epsilon = \left(4\pi \right)^{-\epsilon}\mu^{2\epsilon} e^{\epsilon\gamma_E}$. In anticipation of the more general case needed for vector boson fusion, we have factored the gauge group generators from the integral $I_F$. For the form-factor calculation with only two particles, $\bT_f+\bT_0=0$ since the vertex is a gauge singlet, so that $-\bT_f \cdot \bT_0 = \bT_f^2 = \bT_0^2 = C_F$.

Including the wave-function graphs gives the scalar form factor
%%%
\begin{eqnarray}
F_S &=& \frac{\alpha C_F}{4\pi}\Biggl\{ \frac{3}{\epsilon} + \lQ -4 \lM  +\frac{3}{2}\nn
&&\hspace{-1cm} - \left(2 -\frac{M^2}{\vq^2}\right) \left[ \frac{\pi^2}{6}-\text{Li}_2\left(1-\frac{\vq^2-i0^+}{M^2}\right)\right]
 \Biggr\}\,,
\end{eqnarray} 
%%%
which is gauge invariant. If $\vq^2=0$, as is the case for a VEV ($p_0=0$),
%%%
\begin{eqnarray}
I_F &=& -\frac{1}{\epsilon} + \lM - 1\,,\nn
F_S &=& \frac{\alpha C_F}{4\pi}\Biggl( \frac{3}{\epsilon} -3\lM +\frac{5}{2}\Biggr)\,.
\label{eq15}
\end{eqnarray} 
%%%

%%%
\begin{figure}
\begin{tabular}{ccc}
\includegraphics[width=3cm]{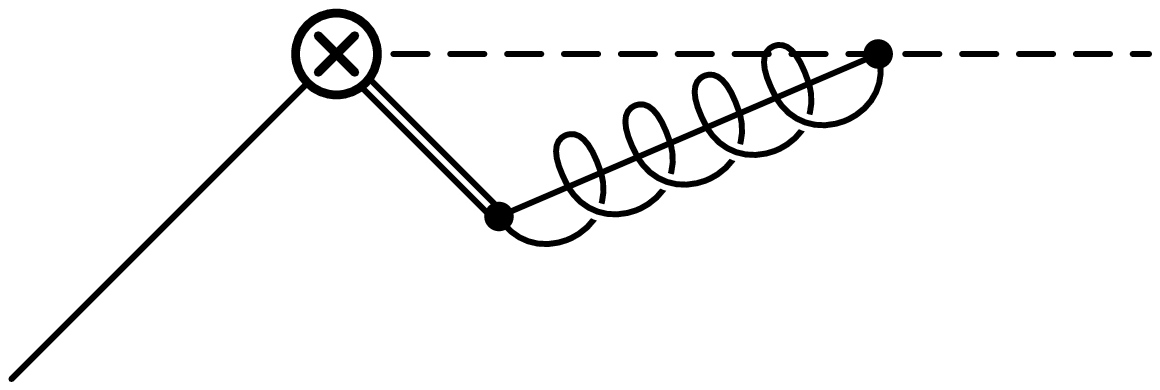} & \qquad &
\includegraphics[width=3cm]{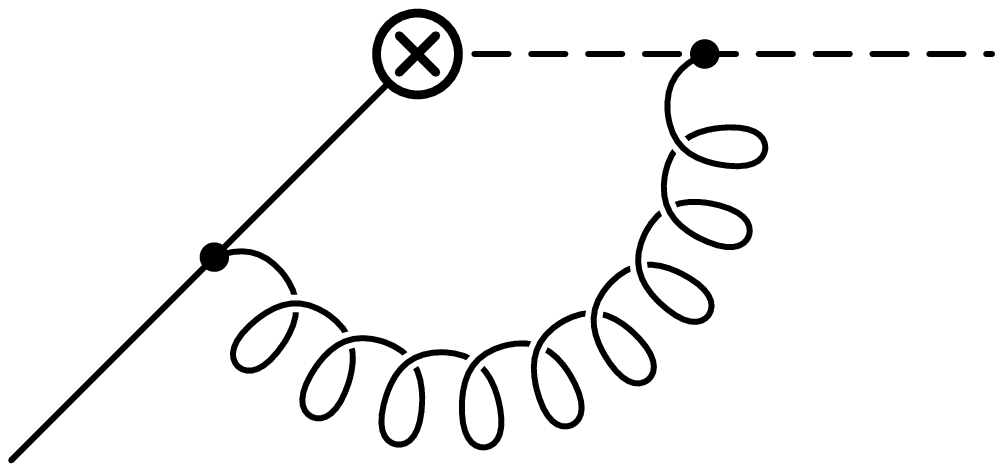} \\
(a) && (b) \\[0.5cm]
\includegraphics[width=3cm]{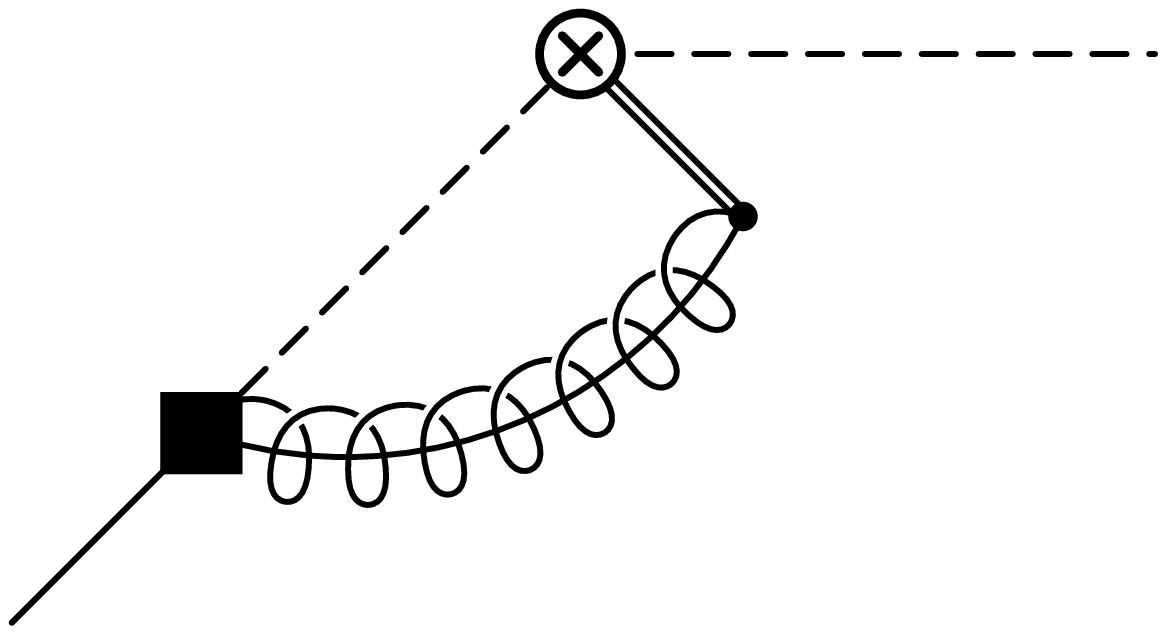} &&
\includegraphics[width=3cm]{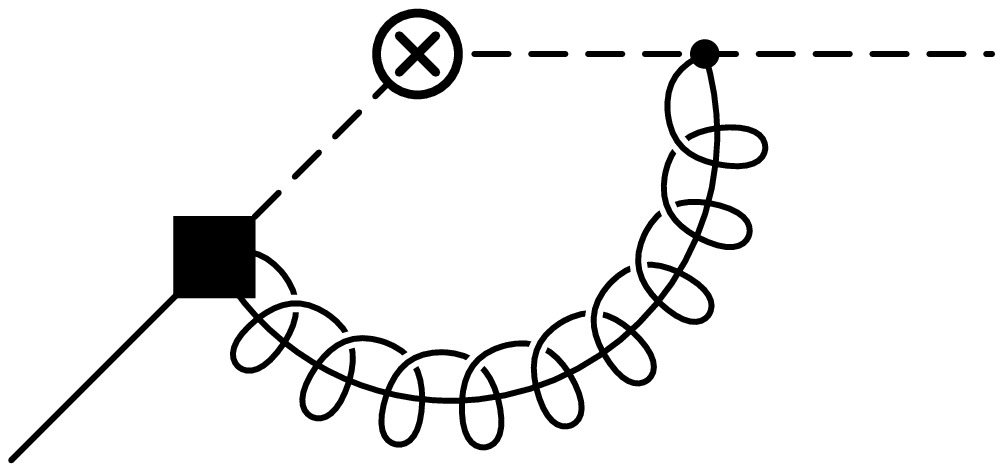} \\
(c) && (d) 
\end{tabular}
\caption{\label{fig:eftloop} One-loop EFT diagrams of the scattering of a scalar field by the external operator $\phi^\dagger \phi$. Graph (a) is the $n$-collinear graph $D_n$, (b) is the soft graph $\widetilde D_s$, and (c) and (d) are graphs $D_1$ and $D_2$ involving the soft-collinear transition vertex. Dashed lines are collinear scalars, and solid lines are soft scalars. The double line denotes a Wilson line.}
\end{figure}
%%%

We will now calculate the graphs in the EFT, regulating the IR using the $\Delta$ regulator~\cite{Chiu:2009yx}. This introduces the regulator parameters $\Delta_{f}$ and $\delta_{f}$ for the collinear field, which are related by
%%%
\begin{eqnarray}
\delta_f &\equiv & \frac{\Delta_f}{\bn \sdt p_f}\,.
\label{dreln}
\end{eqnarray}
%%%
It is not necessary to introduce an IR regulator for the soft particle for computing the zero-bin subtracted diagrams.

The graph Fig.~\ref{fig:eftloop}(a) is the usual $n$-collinear graph where the gauge boson couples to the outgoing field, and to the $n$-collinear Wilson line $W_n$ at the hard vertex. The diagram yields the integral
%%%
\begin{eqnarray} \label{eq17}
D_n &=& \frac{\alpha}{4\pi} \bT_f^2\, I_n\,, \\
I_n &=&-i (4\pi)^2 f_\epsilon \int\! \frac{\rd^d k}{(2\pi)^d} \frac{1}{k^2-M^2}\frac{\bn \sdt (2p_f+k) }{\bn \sdt k\,[(k+p_f)^2-\Delta_f]},\nonumber
\end{eqnarray}
%%%
and the zero-bin subtraction is
%%%
\begin{eqnarray}
I_{n,\zb} &=&-i (4\pi)^2 f_\epsilon \int\! \frac{\rd^d k}{(2\pi)^d} \frac{1}{k^2-M^2}\nn
&& \qquad \times \frac{ 2\bn \sdt p_f }{\bn \sdt k\,[(\bn \sdt p_f)( n \sdt k)-\Delta_f]},
\end{eqnarray}
%%%
which combine to give~\cite{Chiu:2009yx}
%%%
\begin{eqnarray}
\hspace{-3ex} I_n-I_{n,\zb} &=&
\frac{2}{\epsilon^2}-\frac{2}{\epsilon}
\log\frac{\Delta_f}{\mu^2}+\frac{1}{\epsilon}- \log^2 \frac{M^2}{\mu^2}\nn
&&-\left(1-2\log\frac{\Delta_f}{\mu^2}\right)\lM -\frac{\pi^2}{2}+1.
\label{1subs}
\end{eqnarray}
%%%
The zero-bin subtracted collinear graph depends only on the properties of $p_f$, and is independent of the properties of the other particle $p_0$. This is an example of factorization, for which zero-bin subtractions are crucial.

Fig.~\ref{fig:eftloop}(b) is due to soft gauge boson exchange between the soft and collinear lines,
%%%
\begin{eqnarray}
\widetilde D_s &=& \frac{\alpha}{4\pi} \bT_0 \sdt \bT_f\, \widetilde I_s\,, \\
\widetilde I_s &=&i (4\pi)^2 f_\epsilon\! \int \!\!  \frac{{\rm d}^d k}{(2 \pi )^d}\frac{1}{k^2\!-\!M^2}  \frac{1}{n\sdt k-\delta_f}  \frac{n \sdt (2p_0\!+\!k)}{(k\!+\!p_0)^2} .\nonumber
\label{eq73}
\end{eqnarray}
%%%
The ``$\widetilde{ \phantom{i_s}}$" is a reminder that this soft diagram is different from $I_s$ in deep inelastic scattering~\cite{manohar:2003vb}, where the soft gauge boson exchange is between two collinear lines. We evaluate $\widetilde I_s$ by performing the $k^+$ integral by contours, followed by the $k_\perp$ and $k^-$ integrals, yielding\footnote{Some tricky aspects of evaluating SCET integrals using contours are discussed in Appendix~\ref{sec:app}. Alternatively, one can combine the linear and quadratic propagators using the modified Feynman parameter method as in HQET.}
%%%
\begin{eqnarray}
\hspace{-3ex} \widetilde I_s
&=& - \frac{1}{\epsilon} + \lM - 1 - (2+\tau) G(\tau)\,,
\label{eq79}
\end{eqnarray}
%%%
where
%%%
\begin{eqnarray}
\hspace{-4ex}
G(\tau)&=&\left(\frac{1}{\epsilon}-\lM\right)\log \frac{\tau}{1+\tau} - \text{Li}_2\left(\frac{1}{1+\tau}\right),
\label{eq26}
\end{eqnarray}
%%%
and
%%%
\begin{eqnarray}
\tau &=& \frac{\Delta_f}{\vq^2} \,.
\label{eq27}
\end{eqnarray}
%%%
The $\Delta_f$ is a regulator, so the integrals are evaluated with $\Delta_f \to 0$. However, we want to study the vertex correction where $p_0$ is soft, but nonzero, and also the case where $p_0$ is exactly 0. Depending on whether $Q^2 =0$ or $Q^2\not=0$, the $\Delta_f \to 0$ limit corresponds to $\tau \to \infty$ or $\tau \to 0$. We therefore retain the entire $\tau$ dependence, and show that it cancels in the final result. If $Q^2=0$ so that $\tau=\infty$, $\widetilde I_s=0$.

There are also several diagrams involving the soft-collinear transition vertex. Fig.~\ref{fig:eftloop}(c) arises from the gauge boson in the transition vertex coupling to the Wilson line $W_n$ at the hard-scattering vertex,
%%%
\begin{eqnarray}
D_1 &=& \frac{\alpha}{4\pi} \bT_0 \sdt (\bT_0+\bT_f)\, I_1\,, \nn
I_1 &=& -i (4\pi)^2 f_\epsilon \int\! \frac{\rd^d k}{(2\pi)^d} \frac{1}{k^2-M^2}
\frac{1}{(k+p_0)^2} \nn
&=&  \frac{1}{\epsilon}-\lM+1\,.
\end{eqnarray}
%%%
The gauge group factor for this graph is different from the other diagrams. The hard vertex is the operator $\bar \chi_{n_f} \chi_{n_f} = (\xi^\dagger_{n_f}W_{n_f})(W^\dagger_{n_f} \xi_{n_f})= \xi^\dagger_{n_f} \xi_{n_f}$, where the Wilson lines cancel since the fields fields have the same collinear direction. Graph Fig.~\ref{fig:eftloop}(c) is therefore absent in the form-factor calculation. We also want to apply the results of this section to the case where there are many collinear particles at the vertex. In this case, the gauge group factor  for Fig.~\ref{fig:eftloop}(c) is $\bT_0\cdot (\bT_0+\bT_f)$ rather than $\bT_0 \cdot \bT_f$ as for the other graphs in the figure. However, the zero-bin subtraction for graph $D_1$ is
%%%
\begin{eqnarray}
I_{1,\zb} &=& I_1\,,
\end{eqnarray}
%%%
so that the zero-bin subtracted diagram $I_1-I_{1,\zb}$ vanishes. Thus Fig.~\ref{fig:eftloop}(c) does not contribute even in the more general case of processes with several external particles.

Fig.~\ref{fig:eftloop}(d) arises from the gauge boson in the transition vertex coupling directly to the outgoing collinear field and is given by
%%%
\begin{eqnarray}
D_2 &=& \frac{\alpha}{4\pi} \bT_0 \sdt \bT_f\, I_2\,, \nn
I_2 &=&i (4\pi)^2 f_\epsilon \int\! \frac{\rd^d k}{(2\pi)^d} \frac{1}{k^2-M^2}
\frac{1}{(k+p_0)^2} \nn
&&\times \frac{\bn \sdt k\,(2p_f+k)\sdt k - k^2\ \bn \sdt (2p_f+k)}{\bn \sdt k\,[(k+p_f)^2-\Delta_f]}\,.
\end{eqnarray}
%%%
The corresponding zero-bin subtraction is
%%%
\begin{eqnarray}
I_{2,\zb} &=&i (4\pi)^2 f_\epsilon \int\! \frac{\rd^d k}{(2\pi)^d} \frac{1}{k^2-M^2}
\frac{1}{(k+p_0)^2} \nn
&&\qquad \qquad \times \frac{\bn \sdt k \, n \sdt k - 2k^2}{\bn \sdt k\, (n \sdt k-\delta_f)}\,.
\end{eqnarray}
%%%
It is convenient to write
%%%
\begin{eqnarray}
I_2&=& I_n + \overline I_2\,, \qquad
I_{2,\zb}= I_{n,\zb} + \overline I_{2,\zb}\,,
\end{eqnarray}
%%%
where $I_n$ is the $n$-collinear integral of Eq.~(\ref{eq17}), and similarly for the zero-bin subtraction. This gives easier integrals to evaluate,
%%%
\begin{eqnarray}
\overline I_{2}&=& -\frac{1}{\epsilon} -\lQ + 2\lM\nn
&& +\left(2-\frac{M^2}{\vq^2}\right) \left[\frac{\pi^2}{6}-  \text{Li}_2\left(1-\frac{\vq^2}{M^2}\right)\right], \nn
\overline I_{2,\zb}&=& -\frac{1}{\epsilon} + \lM-1-(2+\tau)G(\tau)\,. 
\end{eqnarray}
%%%
On comparing with Eq.~(\ref{eq39}) and Eq.~(\ref{eq79}), we find that $\overline I_2=I_F$, $\overline I_{2,\zb} = \widetilde I_s$. 

\begin{figure}
\includegraphics[width=3cm]{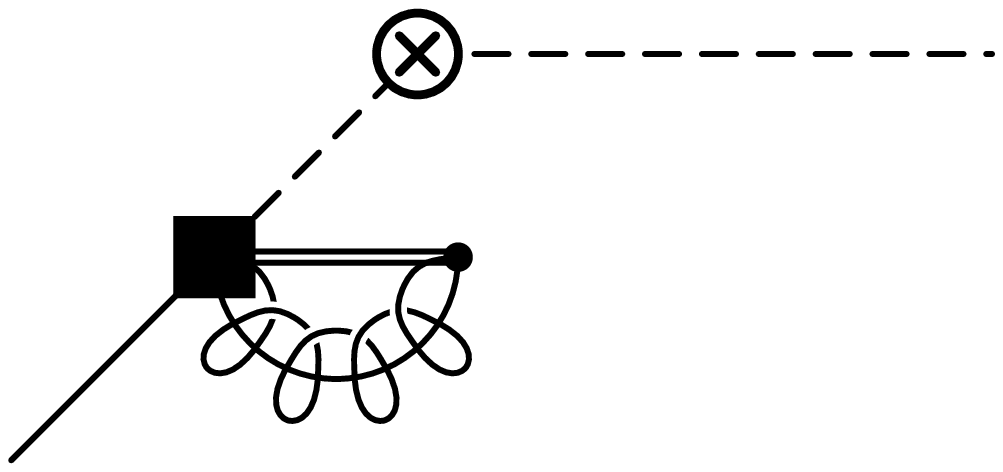} \qquad
\includegraphics[width=3cm]{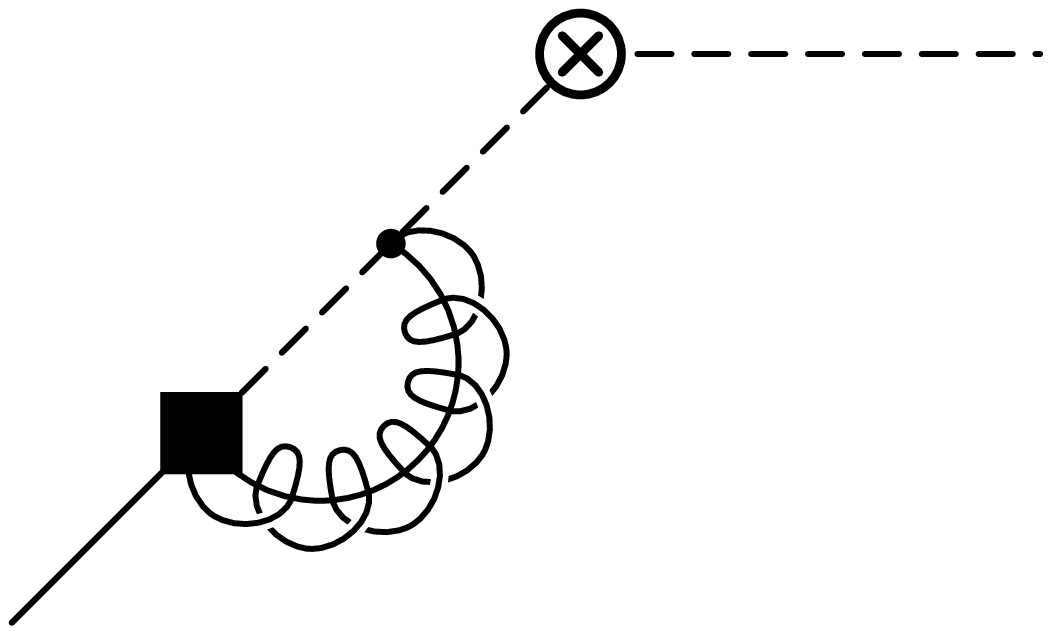}
\caption{\label{fig:eftb} Additional graphs involving the soft-collinear transition vertex, whose contribution vanishes. The dashed lines are collinear fields, and the solid line is a soft field.}
\end{figure}

There are additional graphs involving the soft-collinear transition vertex, given in Fig.~\ref{fig:eftb}.
%%%
%%%
The first graph arises from the gauge boson in the transition vertex coupling to $W_n$ in the transition vertex, and the second from the gauge boson coupling directly to the incoming collinear line. These graphs turn the incoming $p_0$ line into a collinear line, which cannot conserve label momentum since the sums in \eqs{eq7}{eq10} exclude the zero-bin. Their contribution therefore vanishes.

The total vertex correction in the effective theory is the sum of the zero-bin subtracted graphs in Fig.~\ref{fig:oneloop},
%%%
\begin{eqnarray}
  &&\frac{\alpha}{4\pi} [ \bT_f^2 (I_n - I_{n,\zb}) + \bT_0 \sdt \bT_f ( \widetilde I_s + I_2 - I_{2,\zb}) ] \nn
  &&= \frac{\alpha}{4\pi} [ \bT_f \sdt (\bT_0 +\bT_f) (I_n - I_{n,\zb}) + \bT_0 \sdt \bT_f\, I_F ].
\end{eqnarray}
%%%
For the form factor, gauge invariance $\bT_0 + \bT_f=0$ implies that this is equal to the full theory vertex graph $D_F$. The wave-function graphs for the two particles in the full and effective theories also agree, so that the on-shell form factors are equal in the two theories. The entire regulator dependence of the EFT graphs cancels, and zero-bin subtractions~\cite{zerobin} in the EFT theory are necessary for the EFT to correctly reproduce the full theory.

The above computation also shows that the one-loop contribution to the matching for the soft-collinear form factor vanishes since the full and effective theory results agree. This is in contrast to e.g.~deep inelastic scattering, where the incoming and external particles are collinear fermions scattered by the electromagnetic current. There the full and effective theory results differ by a finite matching correction~\cite{manohar:2003vb}
%%%
\begin{eqnarray}
C(Q^2,\mu) &=& \frac{\alpha(\mu) C_F}{4\pi} \biggl(-\log^2 \frac{Q^2}{\mu^2} + 3\lQ + \frac{\pi^2}{6}-8\biggr)\,, \nn
\end{eqnarray}
%%%
which only depends on the short-distance scale $\vq^2$, and is independent of any infrared physics. In our computation, $Q^2$ is no longer a hard scale, so the entire $Q^2$ dependence must be reproduced by the effective theory.

We have explicitly shown the agreement between the full and effective theories for the scalar-scalar form factor with vertex $\phi^\dagger \phi$. One can similarly check agreement for other form factors, such as the scalar-fermion form factor with vertex $\bar \psi \phi$, etc. The computations are very similar to the scalar-scalar case, and will not be given here.

\begin{figure}
\includegraphics[width=3cm]{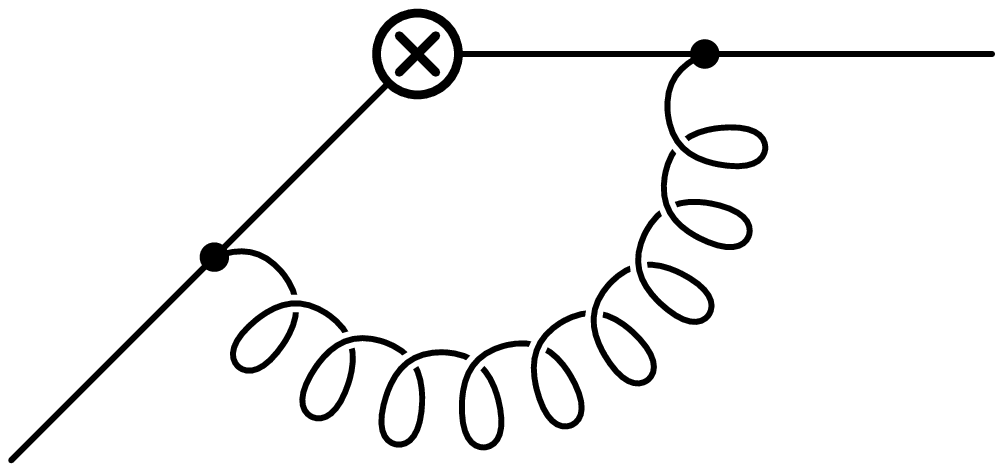}
\includegraphics[width=3cm]{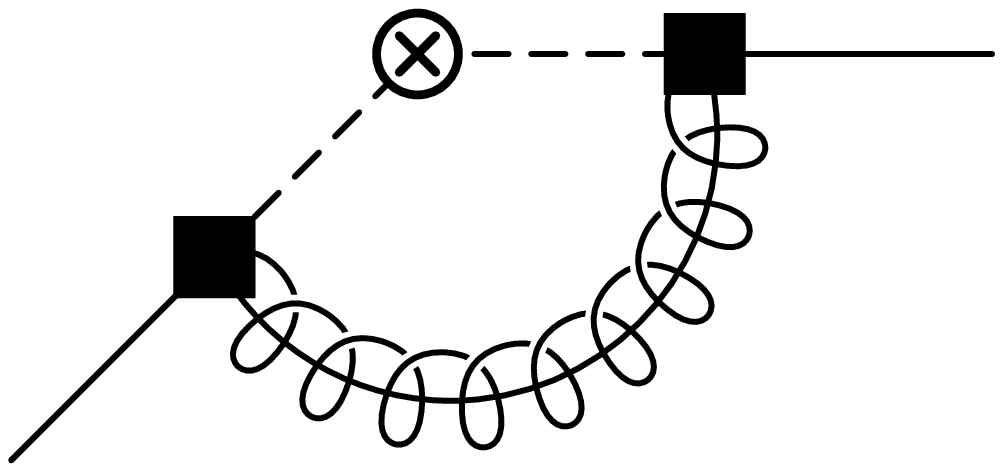}
\caption{\label{fig:3} Vertex graphs when both external particles are soft. The first graph has a soft gauge boson, and the second has a collinear gauge boson. The dashed lines are collinear fields, and the solid lines are soft fields.}
\end{figure}

Lastly, the form factor when both external particles are soft is given by the graphs in Fig.~\ref{fig:3}.
%%%
%%%
The first diagram, with a soft gauge boson loop, reproduces the full theory graph. The soft-collinear transition vertex Eq.~(\ref{eq7}) gives the second graph shown in the figure. The internal particles are $n$-collinear, but the external particles are soft. There is nothing in the kinematics to fix the collinear direction $n$, and the graph does not depend on any external collinear momenta. After zero-bin subtraction, the graph vanishes for any $n$, as it should.

%%%%%%%%%%%%%%%%%%%%%%%%%%%%%%%%%%%%%%%%%%%%%%%%%%%%%%%%%%%%%%%%%%%%%%%%%%%%%%%%
\section{Vector Boson Fusion}\label{sec:vbf}
%%%%%%%%%%%%%%%%%%%%%%%%%%%%%%%%%%%%%%%%%%%%%%%%%%%%%%%%%%%%%%%%%%%%%%%%%%%%%%%%

The previous section treated the one-loop form factor for scalar production by $\phi^\dagger \phi$. We can now apply the formalism to vector boson fusion. We will compute the EFT amplitude, which consists of the running between $\mu_h$ and $\mu_l$, and the low-scale matching. This sums all the large logarithms in the scattering amplitude. The remaining details, such as the high-scale matching and numerical results, will be given in Ref.~\cite{siringo}.

The Higgs part of the vector boson fusion process is described by the operators $\phi^\dagger \phi$ and $\phi^\dagger T^a \phi$, which come from $WW$, $WB$ and $BB$ in the kinetic term for $\phi$. At the high scale, these operators match onto two terms in SCET,
%%%
\begin{eqnarray}
\phi^\dagger \phi &\to&  \left(\phi^\dagger_n W_n\right) \phi_0 + \phi_0^\dagger \left(W_n^\dagger \phi_n\right)\,, \nn
\phi^\dagger T^a \phi &\to&  \left(\phi^\dagger_n W_n\right) T^a \phi_0 + \phi_0^\dagger T^a \left(W_n^\dagger \phi_n\right)\,.
\label{highmatch}
\end{eqnarray}
%%%
The first term on each line has $\phi^\dagger$ replaced by a collinear scalar field $\phi_n^\dagger$, and the $\phi$ field replaced by a soft field $\phi_0$. In the second term, the roles of $\phi$ and $\phi^\dagger$ are reversed. The collinear fields $\left(\phi^\dagger_n W_n\right)$ and $\left(W_n^\dagger \phi_n\right)$ both have momentum equal to $p_h$, the momentum of the Higgs particle. In this section we take all momenta to be incoming, in contrast to the previous section.
The SCET operators that we will need for vector boson fusion are
%%%
\begin{eqnarray}
\op_1 &=& \left[ \left(\phi^\dagger_n W_n\right) \phi_0 + \phi_0^\dagger \left(W_n^\dagger \phi_n\right)
\right] O\,, \nn
\op_2 &=& \left[ \left(\phi^\dagger_n W_n\right) T^a \phi_0 - \phi_0^\dagger T^a \left(W_n^\dagger \phi_n\right) \right] O^a\,, \nn
\op_3 &=& \left[ \left(\phi^\dagger_n W_n\right) \phi_0 - \phi_0^\dagger \left(W_n^\dagger \phi_n\right)
\right] O\,, \nn
\op_4 &=& \left[ \left(\phi^\dagger_n W_n\right) T^a \phi_0 + \phi_0^\dagger T^a \left(W_n^\dagger \phi_n\right)
\right] O^a\,,
\label{simplebasis}
\end{eqnarray}
%%%
where $O$ contains the remaining fermions, which are collinear fields in different directions. (A complete basis for $O$ and $O^a$ is given in Appendix~\ref{sec:basis}.) As is clear from \eq{highmatch}, the operators $\op_2$ and $\op_3$ do not get matched on to at the high scale. However, the running generates these operators and they contribute to the low-scale matching.

At tree level, the low-scale matching amounts to picking out the $h$ component of $\phi_n$ or $\phi^\dagger_n$, and the $v$ component of $\phi_0$ or $\phi^\dagger_0$,
%%%
\begin{eqnarray}
\left(\phi_n^\dagger W_n\right)\! \phi_0 &\to& \frac v2 h_n \,,
\hspace{9.6ex}
\phi_0^\dagger \!\left(W_n^\dagger \phi_n\right) \to \frac v2 h_n\,, \\
\left(\phi_n^\dagger W_n\right)\! T^a \phi_0 &\to& -\frac v4 h_n \delta_{a3} \,,
\hspace{2ex}
\phi_0^\dagger T^a \!\left(W_n^\dagger \phi_n\right) \to -\frac v4 h_n \delta_{a3}\,, \nonumber
\end{eqnarray}
%%%
where $h_n$ creates collinear Higgs particles ($h$ is a real field). Consequently, $\op_2$ and $\op_3$ do not contribute to the low-scale matching at tree level.

The one-loop amplitude can now be computed using the results of the previous section and the analysis in Ref.~\cite{Chiu:2009mg}. We first calculate the anomalous dimension using the unbroken $SU(2) \times U(1)$. Since the spontaneous symmetry breaking is a low energy effect, it does not affect the renormalization. We then repeat the calculation of the anomalous dimension in the broken phase, and determine the low-scale matching. We conclude this section with a brief discussion of the case where one (or both) of the gauge bosons gets produced on shell.

%===============================================================================
\subsection{Unbroken Phase}
%===============================================================================

We start this section by considering a general gauge group, and derive the anomalous dimension for an operator with collinear fields $i=1,r-1$ in distinct collinear directions and a soft scalar field labeled 0. We then specialize to vector boson fusion in the Standard Model and add the contributions from other (non-gauge) interactions.

Adding up the one-loop gauge boson exchange diagrams, we obtain
%%%
\begin{eqnarray} \label{unbr}
 &&\frac{\al}{4\pi} \Big[ \sum_i \bT_i^2 \big(I_n^i - I_{n,\zb}^i \big)
 + \sum_{(ij)} \bT_i \sdt \bT_j I_s^{ij} \label{oneloopres}
 \\
 &&+\, \sum_i \bT_i \sdt \bT_0 \big(I_n^i - I_{n,\zb}^i + I_F^i \big)  \Big] + \sum_i \frac{1}{2} \de R^i + \frac{1}{2} \delta R^\phi
 \,. \nonumber
\end{eqnarray}
%%%
Here the sum on $i$ runs over the collinear fields, but not the soft field. The first term corresponds to the collinear graph in \fig{eftloop}(a) given by $I_n - I_{n,\zb}$. We included the superscript ``$i$" everywhere to account for the different particle types (scalar or fermion). The sum on $(ij)$ in the second term runs over pairs of collinear fields, and describes the contribution $I_s$ from soft gauge boson exchange between the two collinear fields (not to be confused with $\widetilde I_s$). We take the momentum of the soft scalar field $p_0$ to be exactly zero, so the only remaining diagram is \fig{eftloop}(d), which we write as $I_2 = I_n - I_{n,\zb} +I_F$. The final terms are the wave-function contribution $\de R^i$ for the collinear fields and $\de R^\phi$ for the soft scalar field.
Using gauge invariance of the operator, $\bT_0 + \sum_i \bT_i = 0$, we see that the $\Delta$ regulators cancel. 

From the divergent terms in \eq{oneloopres}, we obtain the one-loop anomalous dimension
%%%
\begin{eqnarray}
 \mathbf{\ga}_\op &=& \frac{\al}{4\pi} \bigg\{ 
 \sum_{i \in \phi} \bigg[\bT_i^2 \Big(4 \log \frac{\bn_i \sdt p_i}{\mu} \!-\! 4 \Big) + \bT_i \sdt \bT_0 \Big(4 \log \frac{\bn_i \sdt p_i}{\mu} \Big) \bigg]
 \nn && 
+  \sum_{i \in \psi} \bigg[\bT_i^2 \Big(4 \log \frac{\bn_i \sdt p_i}{\mu} \!-\! 3 \Big) + \bT_i \sdt \bT_0 \Big(4 \log \frac{\bn_i \sdt p_i}{\mu} \!-\! 2 \Big)\bigg] 
\nn &&
- 4 \sum_{(ij)} \bT_i \sdt \bT_j \log \frac{-n_i \sdt n_j - i0}{2} - 2 \bT_0^2 \bigg\}
\,.
\end{eqnarray}
%%%
Here we separated the sum into a sum over collinear scalars ($i \in \phi$) and collinear fermions ($i \in \psi$). We immediately identify the usual collinear and soft functions~\cite{Chiu:2009mg, Chiu:2009ft}, but there are additional terms involving $\bT_i \cdot \bT_0$ or $\bT_0^2$, which need to be included. Define $\tilde \gamma_{\op}$ to be the anomalous dimension for the SCET vector boson fusion operator, given by summing over the soft and collinear functions for all the external particles (i.e.\ fermions and Higgs, but not the VEV). Then the total anomalous dimension is
%%%
\begin{eqnarray}
 \mathbf{\ga}_\op &=&  \tilde \gamma_{\op} + \frac{\al}{4\pi} \bigg[ 
 \bT_h \sdt \bT_0 \Big(4 \log \frac{\bn_h \sdt p_h}{\mu} \Big) - 2 \bT_0^2
 \nn && 
 + \sum_{i \in \psi}\bT_i \sdt \bT_0 \Big(4 \log \frac{\bn_i \sdt p_i}{\mu} \!-\! 2 \Big)\bigg] 
\,.
\label{eq38}
\end{eqnarray}
%%%

\begin{figure}
\includegraphics[width=5cm]{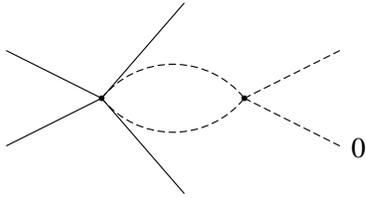}
\caption{\label{fig:2} Radiative correction due to the $\lambda \phi^4$ Higgs self-coupling. The soft field is denoted by $0$. The dashed lines are scalars and the solid lines are fermions.}
\end{figure}

For vector boson fusion in the Standard Model, the Higgs wave-function renormalization receives other contributions, such as from the top Yukawa coupling $y_t$. This is automatically included by using the form in Eq.~(\ref{eq38}) for the anomalous dimension, since the Higgs collinear function defined in Refs.~\cite{Chiu:2009mg, Chiu:2009ft} includes the full Higgs wave-function renormalization. An additional contribution arises from the $\phi$ rescattering graph in Fig.~\ref{fig:2}. 
%%%
%%%
For $\op_2$ and $\op_3$ this vanishes due to the relative minus sign between the terms. The integral is the same for $\op_1$ and $\op_4$, but the coefficient differs: from the contractions of $\phi^\dagger \phi$ and $\phi^\dagger T^a \phi$ with $(\phi^\dagger \phi)^2$, we find a relative factor of 3. 
This leads to an additional contribution to the anomalous dimension
%%%
\begin{eqnarray}
\delta \gamma_\op &=&  \frac{c^\op\lambda}{4\pi^2} \mathbf{\openone} \,.
 \label{garescat}
\end{eqnarray}
%%%
where the coefficient $c^\op = \{3,0,0,1\}$ for the operator $\{\op_1,\op_2,\op_3,\op_4\}$. 

In Refs.~\cite{Chiu:2009mg, Chiu:2009ft}, there was no $\lambda$ contribution to the anomalous dimension for Higgs pair production. This is because Higgs pair production has two energetic Higgs fields moving in different directions, described by collinear SCET fields with different labels. The rescattering of two collinear particles moving in different directions by a $\lambda \phi^4$ coupling is power suppressed. In vector boson fusion, one of the scalars is a soft scalar, and soft-collinear scattering is leading order. 

In Appendix~\ref{sec:basis} a complete basis of SCET operators for vector boson fusion is given, including the action of the $SU(2)$ generators $\bT_i \sdt \bT_0$ in this basis.

%===============================================================================
\subsection{Broken Phase}
%===============================================================================

We will now match onto the broken $SU(2) \times U(1)$ to compute the low-scale matching, and verify that the anomalous dimension computed in the broken and unbroken theories are identical. The diagrams are different in the two phases, because of a change in the gauge-fixing term.

\begin{figure}
\includegraphics[width=5cm]{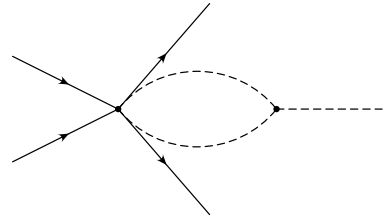}
\caption{\label{fig:5} Radiative correction to Higgs production due to the $\lambda \phi^4$ coupling in the broken phase. The tri-scalar coupling is proportional to $\lambda v$. The dashed lines are scalars and the solid lines are fermions, which are all collinear.}
\end{figure}

At tree level, the $\phi^\dagger \phi$ and $\phi^\dagger T^a \phi$ part of \eq{simplebasis} turn into
%%%
\begin{eqnarray}
\op_1 &\to& \Big[(v + h_0) h_n + \vp_0^a \vp_n^a\Big]\, O\,, \nn
\op_2 &\to& \frac{i}{2} \Big[-(v + h_0)\vp_n^a + \vp_0^a h_n + \eps^{abc} \vp_0^b \vp_n^c\Big] \, O^a,\hspace{2ex}
\label{brhigh}
\end{eqnarray}
%%%
where $\vp^a$ are the unphysical Goldstone bosons. The corresponding expressions for $\op_3$ and $\op_4$ can be obtained from those for $\op_1$ and $\op_2$, by replacing
%%%
\begin{eqnarray}
 h_n \lra i\vp_n^3\,, \quad \vp_n^1 \lra i \vp_n^2\,.
 \label{op34repl}
\end{eqnarray}
%%%
The terms in \eq{brhigh} with more than a single $h_n$ or $\vp_n^a$ can contribute to vector boson fusion through e.g.~the rescattering graph in \fig{5}, which is the analog of \fig{2} in the broken phase. 
%%%
%%%
From the calculation in the unbroken phase, we know that all the terms in \eq{brhigh} for the same operator have the same renormalization and thus stay together. We can therefore restrict ourselves to graphs with one external collinear scalar to crosscheck the anomalous dimension. Only the graphs with one external Higgs boson contribute to the low-scale matching. In the low-scale matching we get a dependence on the masses of particles for the first time, and also need to take into account that only $W$ and $Z$ (but not $\gamma$) get integrated out. The following identity will be useful:
%%%
\begin{eqnarray}
  &&\al_2 \bT_i \sdt \bT_j + \al_1 \mathbf{Y}_i \mathbf{Y}_j  \\
  &&= \al_W (\bT_i^+ \bT_j^- + \bT_i^- \bT_j^+) + \al_Z \bT_i^Z \bT_j^Z + \al_\text{em} \mathbf{Q}_i \mathbf{Q}_j \nn
  &&= \al_2 (\bT_i \sdt \bT_j - \bT_i^3 \bT_j^3) + (\al_2 + \al_1) \bT_i^Z \bT_j^Z + \al_\text{em} \mathbf{Q}_i \mathbf{Q}_j \,, \nonumber
 \label{couplbr}
\end{eqnarray}
%%%
where $\bT^\pm = (\bT^1 \pm i \bT^2)/\sqrt{2}$, $\bT^Z = \bT^3 - \sin^2 \theta_W \mathbf{Q}$, $\mathbf{Y}$ is the hypercharge and $\mathbf{Q}$ is the charge operator.

\begin{figure}
\includegraphics[width=3cm]{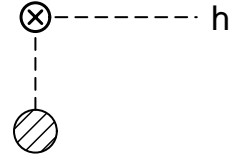}
\caption{\label{fig:12} Tadpole correction to Higgs production. The tadpole amplitude $\Gamma^h$ is denoted by the shaded blob.}
\end{figure}

We will restrict ourselves to studying contributions that involve the Higgs sector. The diagrams are: soft gauge boson exchange between a scalar and a fermion akin to \fig{oneloop}, the collinear graph for a scalar in \fig{eftloop}(a), the scalar wave-function diagram, the rescattering graph in \fig{5}, and the tadpole graph in \fig{12}. 
%%%
%%%
In the above we wrote ``scalar" to allow for both an external Higgs or Goldstone boson. In the low-scale matching we only need the Higgs contribution.

The soft gauge boson exchange can be written in the same form as in the unbroken phase in \eq{oneloopres}, with the understanding that
%%%
\begin{eqnarray}
  \hspace{-3ex}
  \bT_h^a h_n = -\frac{i}{2} \vp_n^a\,, \quad 
  \bT_h^a \vp_n^b = \frac{i}{2} (\de^{ab} h_n + \eps^{abc} \vp_n^c)
  \,, \label{brgen}
\end{eqnarray}
%%%
subject to the replacements in \eq{op34repl} for $\op_3$ and $\op_4$.
These expressions can be obtained using the correspondence with the unbroken phase, e.g. 
%%%
\begin{eqnarray}
  \bT_h^a h_n &\to& \frac{1}{v} \bT_h^a \big(\phi_n^\dagger \phi_0 + \phi_0^\dagger \phi_n\big) \nn
  &=& \frac{1}{v} \big(\phi_n^\dagger T^a \phi_0 + \phi_0^\dagger T^a \phi_n\big)  
  \to -\frac{i}{2} \varphi_n^a\,.
\end{eqnarray}
%%%
We also need to take the mass of the Goldstone boson $M_\vp \neq M_h$ into account in the loop integral $I_s$. This leads to the replacement $\De \to \De + M_\text{in}^2 - M_\text{ex}^2$ for the $\De$ regulator of the scalar field, where $M_\text{in}$ is the mass of the scalar in the loop and $M_\text{ex}$ is the mass of the external scalar. Since the $\De$ regulators will cancel out, this shift is irrelevant.

In the collinear graph, a gauge boson is emitted from the Wilson line and reabsorbed later. We would therefore need to include the Wilson line contribution in \eq{brhigh}. This diagram can again be described by the same expression as in the broken phase, using \eq{brgen} and replacing $\De \to \De + M_\text{in}^2 - M_\text{ex}^2$ in the $\De$ regulator of the scalar particle. In the low-scale matching, where the external particle is the Higgs, the massive collinear function has an additional term compared to the massless case in \eq{1subs} 
%%%
\begin{eqnarray}
 I_n^h - I_{n,\zb}^h &=& \big( I_n - I_{n,\zb} \text{ with } \De_h \to \De_h + M_\vp^2 - M_h^2 \big) \nn 
 && + f_S\Big(\frac{M_h^2}{M^2}, \frac{\De_h + M_\vp^2}{M^2}\Big)
\end{eqnarray}
%%%
The function $f_S$ was defined in Appendix B of Ref.~\cite{Chiu:2008vv} 
%%%
\begin{eqnarray}
  f_S(w,z) = \int_0^1 {\rm d} x\, \frac{2-x}{x} \log \frac{1 - x + zx - wx(1-x)}{1-x}
  \,. \nn
\end{eqnarray}
%%%
Since we can set $\De_h = 0$ and $M_\vp = M$, we need
%%%
\begin{eqnarray}
  f_S(w,1) &=& 1 + \frac{\pi^2}{3} - 2 \sqrt{\frac{w-4}{w}} \tanh^{-1} \sqrt{\frac{w}{w-4}} \\
  &&- 2 \text{Li}_2 \frac{w \!+\!\sqrt{(w\!-\!4)w}}{2} - 2 \text{Li}_2 \frac{w \!-\! \sqrt{(w \!-\! 4)w}}{2}
  \,. \nonumber
\end{eqnarray}
%%%

As was discussed at the end of section \sec{eft}, we do not get the $\mathfrak{R}_0$ contribution to the Wilson line in the broken phase. This leads to the replacement $\bT_i^2 \to \bT_i \cdot (\bT_i + \bT_0)$ in \eq{unbr} for the contribution from the collinear graph. Here $\bT_0$ acts like
%%%
\begin{eqnarray}
  \hspace{-3ex}
  \bT_0^a h_n = \frac{i}{2} \vp_n^a\,, \quad 
  \bT_0^a \vp_n^b = \frac{i}{2} (-\de^{ab} h_n + \eps^{abc} \vp_n^c)
  \,, \label{brgen2}
\end{eqnarray}
%%%
with the usual replacements in \eq{op34repl} for $\op_3$ and $\op_4$.
The diagram in \fig{eftloop}(d) is absent in the broken phase, but gets reproduced by this $\bT_i^2 \to \bT_i \cdot (\bT_i + \bT_0)$ replacement.

The divergent part of the one-loop Higgs wave function, which enters the wave-function renormalization, is the same as in the unbroken phase. The finite part will also be denoted by $\de R^h$. Specifically, for the finite part, $R^h = 1+ \de R^h$ is the residue of the pole in the propagator, $S \sim iR^h/(p^2 - M_h^2)$.

Next, we consider the contribution through the rescattering diagram of the terms in \eq{brhigh}. Writing out
%%%
\begin{eqnarray}
  (\phi^\dagger \phi)^2 = v(h^3 + h \vp^a \vp^a) + \dots\,,
\end{eqnarray}
%%%
we note that diagrams involving the $h^3$ interaction gain a factor of 3 compared to $h\vp^a\vp^a$ from symmetry. We reproduce the same results as in the unbroken phase: there is no contribution from $\op_2$ and $\op_3$ and there is a relative factor of 3 between the divergent contribution for $\op_1$ and $\op_4$. The finite terms, which enter in the low-scale matching at  $\mu_l$, are given by
%%%
\begin{eqnarray}
  D_\lambda^\text{fin} &= &-\frac{\lambda}{16\pi^2}\sum_j \eta_j^\op I_\lambda^j \,,\\
  I_\lambda^j &=& 2 \!-\! \log \frac{M_j^2}{\mu_l^2}
  \!-\! 2\sqrt{\frac{M_h^2 \!-\! 4M_j^2}{M_h^2}}\tanh^{-1} \sqrt{\frac{M_h^2}{M_h^2 \!-\! 4M_j^2}}\,. \nonumber
\end{eqnarray}
%%%
Here, the sum over $j$ is over $j=h,Z,W$, with combinatorial weight $\eta_h^\op= \{3,0,0,3\}$, $\eta_W^\op=\{2,0,0,-2\}$, $\eta_Z^\op=\{1,0,0,1\}$ for $\{\op_1,\op_2,\op_3,\op_4\}$. For the anomalous dimension, mass effects are unimportant, which is why we only needed $c^\op = (\eta_h^\op + \eta_W^\op + \eta_Z^\op)/2$.

The tadpole graph in \fig{12} only attaches to $h_0$, and thus contributes universally to all the operators. The scalar one-point function will be denoted by $\Gamma^h$, which we use for both the finite and divergent pieces, depending on the context. Since we are working in terms of operators with one external scalar, we need to remove the VEV and include a 0-momentum propagator to attach the tadpole, leading to $\Gamma^h/(M_h^2 v)$.

Adding up all the diagrams
%%%
\begin{eqnarray}  \label{oneloopbr}
 &&\frac{\al}{4\pi} \Big[ 
 \sum_i \bT_i \sdt (\bT_i + \bT_0) \big(I_n^i - I_{n,\zb}^i \big)
 + \sum_{(ij)} \bT_i \sdt \bT_j I_s^{ij}  \big]
 \nn
 &&
 + \sum_i \frac{1}{2} \de R^i  + D_\lambda + \frac{\Gamma^h}{M_h^2 v}
 \,,
\end{eqnarray}
%%%
where the sum over the $SU(2)$ and $U(1)$ gauge group is implicit. We can use gauge invariance of the operator, $\bT_0 + \sum_i \bT_i = 0$, with the action of $\bT_h$ and $\bT_0$ defined in \eqs{brgen}{brgen2}, to show that the $\De$ regulators cancel. 

We will now compare the divergent terms in \eq{oneloopbr} to the result in the unbroken phase in \eq{oneloopres} [plus the rescattering contribution in \eq{garescat}].
The difference with the unbroken phase amounts to one less Higgs wave-function renormalization, an additional tadpole contribution and no $I_F^\psi = I_F^\phi$ with $\bT_i \cdot \bT_0$ color structure:
%%%
\begin{eqnarray}
 \frac{\Gamma^h}{M_h^2 v} - \frac{1}{2} \de R^h - \frac{\al C_F}{4\pi} I_F = - \de Z_v
 \,.
\end{eqnarray}
%%%
This is equal to the renormalization of the VEV. Since we are now evaluating the VEV at the hard scale, this additional piece in the anomalous dimension converts $v(\mu_h) \to v(\mu_l)$, to give precisely the same result as in the unbroken computation.

%===============================================================================
\subsection{Low-Scale Matching}
%===============================================================================

The last piece of the computation is the low-scale matching at $\mu_l$, where the $W$ and $Z$ are integrated out, and one matches onto a theory with photons and gluons. For the low-scale matching we need the finite part of \eq{oneloopbr}, taking into account that only the $W$ and $Z$ get integrated out [see \eq{couplbr}] and including the effect of the masses.
We find that
\begin{widetext}
%%%
\begin{eqnarray} \label{lowmatch}
\mathbf{D}\op_1 &=& vh_n \bigg[ \bigg( \frac 12 \delta R^h  
- \frac{\lambda}{16\pi^2} \sum_j \eta_j^1 I_\lambda^j+ \frac{\Gamma^h}{M_h^2 v}\bigg) O + \mathbf{D} O \bigg], \nn
\mathbf{D}\op_2 &=& \frac{vh_n}{4} \bigg[\frac{\al_W}{4\pi} \sum_i (I_s^{ih} - I_n^i + I_{n,\zb}^i ) (\bT_i^1 O^1 + \bT_i^2 O^2)  
 + \frac{\al_Z}{4\pi} \sum_i (I_s^{ih} - I_n^i + I_{n,\zb}^i ) \bT_i^Z O^3 \bigg], \nn
\mathbf{D}\op_3 &=& -\frac{vh_n}{2} \bigg[\frac{\al_Z}{4\pi} \sum_i (I_s^{ih} - I_n^i + I_{n,\zb}^i ) \bT_i^Z O \bigg] \,,\nn
\mathbf{D}\op_4 &=& -\frac{vh_n}{2} \bigg\{\bigg[\frac{\al_W}{4\pi} (I_n^h \!-\! I_{n,\zb}^h)  + \frac 12 \delta R^h - \frac{\lambda}{16\pi^2} \sum_j \eta_j^4 I_\lambda^j 
 + \frac{\Gamma^h}{M_h^2 v}\bigg] O^3 + \frac{\al_W}{4\pi} 
\sum_i (I_s^{ih} \!+\! I_n^i \!-\! I_{n,\zb}^i) \frac{1}{2} i\eps^{ab3} \bT_i^a O^b 
+ \mathbf{D} O^3 \bigg\}\,, \nn
\end{eqnarray}
%%%
where the gauge boson mass is $M_W$ ($M_Z$) in the terms involving $\al_W$ ($\al_Z$). The sum $i$ runs over all the fermions in O and $j$ runs over $h, W, Z$. The $\mathbf{D}$ on the right-hand side of this equation denotes the low-scale matching for the fermions. This is given in terms of the collinear and soft functions of Refs.~\cite{Chiu:2009mg,Chiu:2009ft} by reshuffling terms between $I_n^i - I_{n,\zb}^i$ and $I_s^{ij}$, as discussed there. A consequence of this reshuffling is that
%%%
\begin{eqnarray}
 I_n^i - I_{n,\zb}^i &\to& \bigg(2 \log \frac{\bn_i \sdt p_i}{\mu} \log \frac{M^2}{\mu^2} - \frac{1}{2} \log^2 \frac{M^2}{\mu^2} - 2\log \frac{M^2}{\mu^2}  
 - \frac{5\pi^2}{12} + 2\bigg)\,, \nn
 I_n^h - I_{n,\zb}^h &\to& \bigg[2 \log \frac{\bn_h \sdt p_h}{\mu} \log \frac{M^2}{\mu^2} - \frac{1}{2} \log^2 \frac{M^2}{\mu^2} - \log \frac{M^2}{\mu^2}  
 - \frac{5\pi^2}{12} + 1 + f_S\Big(\frac{M_h^2}{M^2}, 1\Big) \bigg]\,, \nn
 I_s^{ih} &\to& -2 \log \frac{-n_i \sdt n_h -i0}{2} \log \frac{M^2}{\mu^2}\,,
\end{eqnarray}
%%%
\end{widetext}
in the above equation. For the operators $\op_2$ and $\op_3$, all the terms that do not depend on the particle $i$ actually cancel out in the sum on $i$. For $\op_4$ a similar cancellation comes about through
%%%
\begin{eqnarray}
 \sum_i \frac{1}{2} i \eps^{ab3} \bT_i^a O^b = - O^3
 \,.
\end{eqnarray}
%%%

The expression Eq.~(\ref{lowmatch}) has been written in a compact form. Gauge operators such as $\mathbf{T}_i^Z$ act differently on $u$ and $d$-type quarks, since they have different $SU(2) \times U(1)$ charges. The explicit expression can be readily computed from Eq.~(\ref{lowmatch}) by expanding out the fermion operators in Appendix~\ref{sec:basis} into individual fermion components. The low-scale matching will be described in full detail in Ref.~\cite{siringo}. 

The total radiative correction can now be obtained from Eqs.~(\ref{eq38}), (\ref{garescat}) and (\ref{lowmatch}). One can run the Higgs production operator in the theory below $\mu_l$. Since the Higgs is $SU(3) \times U(1)$ neutral, the running below $\mu_l$ is identical to that of the operator $O$ or $O^a$. 

%===============================================================================
\subsection{On-Shell Gauge Bosons}
%===============================================================================

The discussion so far has been for the case where the two gauge bosons in Fig.~\ref{fig:fusion} are off shell, so that the EFT amplitude reduces to the local operator in Fig.~\ref{fig:fusioneft}. However, this approach can also be used for more general kinematics. If one gauge boson is off shell and the other is close to on shell, then the off-shell gauge boson can be integrated out, and the vector boson fusion amplitude reduces to a $\bar q q W \phi^\dagger \phi$ interaction, multiplied by the amplitude for on-shell $W$ production (and similarly for the $Z$ boson). The radiative correction to the $\bar q q W \phi^\dagger \phi$ amplitude can be computed using the methods of this paper. The on-shell $W$ production amplitude $q \to q W$ has no large kinematic scales, and so does not contain any large electroweak logarithms. SCET does not simplify the computation of this amplitude as there is no hard scale to integrate out, and the effective field theory result is identical to the full theory computation.

If both $W$ bosons are close to on shell, then the $WW \to H$ amplitude no longer has a large scale, and SCET simply reproduces the full theory for this amplitude.
The computations of $WW \to H$ and  $q \to q W$ are far simpler than those for the full vector boson fusion amplitude, since each amplitude involves only three external on-shell fields, and there are no free kinematic variables.

In Ref.~\cite{Chiu:2009mg}, it was shown that even though SCET neglects $M_Z^2/s$ power corrections, it still reproduces the full amplitude for fermion pair production to 2\% accuracy even at $Q^2=M_Z^2$. Thus using SCET for vector boson fusion, even for gauge bosons close to on shell, should still be a reliable approximation.

%%%%%%%%%%%%%%%%%%%%%%%%%%%%%%%%%%%%%%%%%%%%%%%%%%%%%%%%%%%%%%%%%%%%%%%%%%%%%%%%
\section{Conclusions}
%%%%%%%%%%%%%%%%%%%%%%%%%%%%%%%%%%%%%%%%%%%%%%%%%%%%%%%%%%%%%%%%%%%%%%%%%%%%%%%%

Electroweak radiative corrections to Higgs production via vector boson fusion were computed using SCET. The amplitude is proportional to the gauge symmetry breaking Higgs VEV, so standard resummation methods do not apply. By contrast, standard resummation methods may be used for the QCD corrections, since they receive no contribution from the Higgs sector. 

In the unbroken phase of the gauge group, the VEV is treated as an external soft field. The formalism for treating scattering with both energetic collinear fields and soft fields was developed. There is a new scalar soft-collinear transition vertex, which is part of the subleading SCET Lagrangian and contributes to the amplitude. In the broken phase of the gauge theory, there are no emissions from the line corresponding to the VEV, the contribution necessary to build up collinear Wilson lines is absent too. A subtlety when switching to the broken phase of the gauge group in SCET, is that the soft field changes its power counting when it attains a VEV.

The final result is given in Eqs.~(\ref{eq38}), (\ref{garescat}) and (\ref{lowmatch}). The $\lambda\phi^4$ coupling enters the final answer and is not power suppressed. It gives a numerically significant ($\sim -10\%$) contribution to the cross-section. This is in contrast to processes such as Higgs pair production, where the $\lambda \phi^4$ interaction was subleading in the power counting, and could be neglected. Thus an accurate measurement of the vector boson fusion process might be a way of determining $\lambda$. Detailed numerical results will be presented in a future publication~\cite{siringo}.

%%%%%%%%%%%%%%%%%%%%%%%%%%%%%%%%%%%%%%%%%%%%%%%%%%%%%%%%%%%%%%%%%%%%%%%%%%%%%%%%
\acknowledgments
%%%%%%%%%%%%%%%%%%%%%%%%%%%%%%%%%%%%%%%%%%%%%%%%%%%%%%%%%%%%%%%%%%%%%%%%%%%%%%%%

We would like to thank F.~Siringo for helpful discussions and feedback on the manuscript.
This work was supported in part by DOE grant DE-FG02-90ER40546.

\begin{appendix}

%%%%%%%%%%%%%%%%%%%%%%%%%%%%%%%%%%%%%%%%%%%%%%%%%%%%%%%%%%%%%%%%%%%%%%%%%%%%%%%%
\section{SCET Feynman Integrals}\label{sec:app}
%%%%%%%%%%%%%%%%%%%%%%%%%%%%%%%%%%%%%%%%%%%%%%%%%%%%%%%%%%%%%%%%%%%%%%%%%%%%%%%%

One has to be careful in evaluating SCET Feynman integrals by contours. A simple example is the tadpole graph
%%%
\begin{eqnarray}
I &=&-i g^2 C_F f_\epsilon \int\! \frac{\rd^d k}{(2\pi)^d} \frac{1}{k^2-M^2}\nn
&=& \frac{\alpha C_F}{4\pi}  \left(\frac{1}{\epsilon}-\lM+1 \right) M^2\,,
\end{eqnarray}
%%%
which is nonzero. Evaluating the graph by first doing the $k^+$ integral by contours gives zero. The problem is that the integrand falls like $1/k^+$, so the contour at infinity does not vanish and the method of residues cannot be used for the $k^+$ integral. A more insidious example is the integral
%%%
\begin{eqnarray}
I &=& -i g^2 C_F f_\epsilon\int\! \frac{\rd^d k}{(2\pi)^d} \frac{1}{k^2-M^2}
\frac{1}{k^2} \nn
&=& \frac{\alpha C_F}{4\pi}  \left(\frac{1}{\epsilon}-\lM+1 \right)\,.
\end{eqnarray}
%%%
For large $k^+$, the integrand falls like $1/(k^+)^2$, so the contour at infinity does not contribute. Evaluating the $k^+$ integral by contours gives zero, since the two $k^+$ poles are on the same side of the axis for any value of $k^-$. The flaw in the method is that for $k^-=0$, the integrand does not depend on $k^+$, so the contour integration fails at this value of $k^-$, and there is a $\delta(k^-)$ term that needs to be included. Regulate the integral by introducing a $p^-$,
%%%
\begin{eqnarray}
I & \to & -i g^2 C_F f_\epsilon\int\! \frac{\rd^d k}{(2\pi)^d} \frac{1}{k^2-M^2}
\frac{1}{k^+(k^-+p^-)- \vec k_\perp^{\,2}}\,,\nn
\end{eqnarray}
%%%
so that both $k^+$ coefficients in the denominator do not simultaneously vanish. Then one can evaluate the $k^+$ integral by contours, followed by the $k_\perp$ and $k^-$ integrals, to get the correct result as $p^- \to 0$. The entire contribution to the integral arises from $-p^- \leq k^- \leq 0$, which gives the $\delta(k^-)$ contribution in the limit $p^- \to 0$.

%%%%%%%%%%%%%%%%%%%%%%%%%%%%%%%%%%%%%%%%%%%%%%%%%%%%%%%%%%%%%%%%%%%%%%%%%%%%%%%%
\section{Operator Basis}\label{sec:basis}
%%%%%%%%%%%%%%%%%%%%%%%%%%%%%%%%%%%%%%%%%%%%%%%%%%%%%%%%%%%%%%%%%%%%%%%%%%%%%%%%

We complete \eq{simplebasis} by listing a basis for the $O$ and $O^a$, containing the fermion fields. For Higgs production through vector boson fusion, the $O$ and $O^a$ needed are
%%%
\begin{align}
 O_A &= \bar \Psi_3 \ga^\mu T^a \Psi_1 \bar \Psi_4 \ga_\mu T^a \Psi_2
 \,, \nn 
 O_B &= C_F \bar \Psi_3 \ga^\mu \Psi_1 \bar \Psi_4 \ga_\mu \Psi_2 
 \,, \nn 
 O^a_A &= \bar \Psi_3 \ga^\mu T^a \Psi_1 \bar \Psi_4 \ga_\mu \Psi_2 
 \,, \nn  
 O^a_B &= \bar \Psi_3 \ga^\mu \Psi_1 \bar \Psi_4 \ga_\mu T^a \Psi_2
 \,, \nn 
 O^a_C &= i \eps^{abc}\, \bar \Psi_3 \ga^\mu T^b \Psi_1 \bar \Psi_4 \ga_\mu T^c\Psi_2
\,.
\end{align}
%%%
The subscripts distinguish the (possibly) different fermion fields.
We use the notation $\op_{1A}$ for $\op_1$ with $O_A$, etc.

We now work out the action of $\bT_i \cdot \bT_0$, $\bT_i \cdot \bT_h$ and $\bT_h \cdot \bT_0$ for SU(2). Under $SU(2)$, the renormalization group evolution only mixes these ten operators within the subsets $\{\op_{1A}$, $\op_{1B}$, $\op_{2A}$, $\op_{2B}$, $\op_{2C}\}$ and $\{\op_{3A}$, $\op_{3B}$, $\op_{4A}$, $\op_{4B}$, $\op_{4C}\}$. In fact, for these bases the $\bT_i \cdot \bT_0$, $\bT_i \cdot \bT_h$ and $\bT_h \cdot \bT_0$ have identical expressions for both subsets. Using the notation
%%%
\begin{equation}
 \bT_i \sdt \bT_j \op_k = \sum_m \op_m (\bT_i \sdt \bT_j)_{mk}
 \,,
\end{equation}
%%%
we find
%%%
\begin{align}
 \bT_1 \sdt \bT_0 &= \frac{1}{4}
 \begin{pmatrix}
  0 & 0 & 0 & 1 & 1\\
  0 & 0 & 1 & 0 & 0\\
  0 & 3 & -2 & 0 & 0\\
  1 & 0 & 0 & 0 & -1\\
  2 & 0 & 0 & -2 & -1
 \end{pmatrix}
 \,, \nn
 \bT_2 \sdt \bT_0 &= \frac{1}{4}
 \begin{pmatrix}
  0 & 0 & 1 & 0 & -1\\
  0 & 0 & 0 & 1 & 0\\
  1 & 0 & 0 & 0 & 1 \\
  0 & 3 & 0 & -2 & 0 \\
  -2 & 0 & 2 & 0 & -1
 \end{pmatrix}
 \,, \nn
  \bT_3 \sdt \bT_0 &= \frac{1}{4}
 \begin{pmatrix}
  0 & 0 & 0 & -1 & 1\\
  0 & 0 & -1 & 0 & 0\\
  0 & -3 & -2 & 0 & 0\\
  -1 & 0 & 0 & 0 & 1\\
  2 & 0 & 0 & 2 & -1
 \end{pmatrix}
 \,,\nn \nonumber
 \end{align}
 \begin{align}
 \bT_4 \sdt \bT_0 &= \frac{1}{4}
 \begin{pmatrix}
  0 & 0 & -1 & 0 & -1\\
  0 & 0 & 0 & -1 & 0\\
  -1 & 0 & 0 & 0 & -1 \\
  0 & -3 & 0 & -2 & 0 \\
  -2 & 0 & -2 & 0 & -1
 \end{pmatrix}
 \,, \nn
 \bT_h \sdt \bT_0 &= \frac{1}{4}
 \begin{pmatrix}
  -3 & 0 & 0 & 0 & 0 \\
  0 & -3 & 0 & 0 & 0\\
  0 & 0 & 1 & 0 & 0\\
  0 & 0 & 0 & 1 & 0\\
  0 & 0 & 0 & 0 & 1
 \end{pmatrix}
 \,.
\end{align}
%%%
Here we used $C_A = N = 2$, $C_F = \frac{3}{4}$, and the $SU(2)$ group theory identities
%%%
\begin{align}
  & T^a T^b = \frac{1}{4} \de^{ab} + \frac{i}{2} \eps^{abc} t^c
  \,,
  & & \eps^{abc} t^a t^b = i t^c
  \,, \\
  & \eps^{abc} \eps^{abd} = 2 \de^{cd}
  \,,
  & &\eps^{abc} \eps^{ade} \eps^{bdf} = \eps^{cef}
  \,. \nonumber
\end{align}
%%%
The matrix expressions for $\bT_i \sdt \bT_h$ in this basis can be simply obtained from the above expressions for $\bT_i \sdt \bT_0$ by using
%%%
\begin{eqnarray}
  &&\bT_i \sdt \bT_h = C\, \bT_i \sdt \bT_0\, C
  \,, \nn
  && C =  \begin{pmatrix}
  1 & & & & \\
   & 1 & & & \\
   & & -1 & & \\
   & & & -1 & \\
   & & & & -1
  \end{pmatrix}\!.
\end{eqnarray}
%%%
The action of the remaining generators, as well as complete expressions for the anomalous dimension, will be given in Ref.~\cite{siringo}.

\end{appendix}

\bibliography{vbf}

\end{document}